# Effects of Microwave Irradiation on Multiwalled Carbon Nanotubes of Different Diameters

Parker Adamson and Scott Williams*

Department of Physics, Angelo State University, ASU Station #10904 San Angelo, Texas 76909, USA

We have studied the visible and infrared radiation emitted by multiwalled carbon nanotubes of different diameters when exposed to 2.45 GHz microwaves. A comparison of the spectra suggests that multiwalled carbon nanotubes (MWCNTs) with larger diameters emit radiation of greater intensity than those with smaller diameters. Furthermore, the MWCNTs continued to emit visible and infrared radiation over the course of several microwave-irradiation cycles, with no degradation in the intensity of the emitted radiation. A comparison of Raman D- to G-band peak-intensity ratios revealed that microwave-irradiation did not significantly impact the MWCNTs' defect densities. The results of our experiments suggest that MWCNTs may have the potential for use in lighting technologies, and that ohmic heating caused by the polarization of the MWCNTs in the microwave field is likely responsible for the observed emissions of visible and infrared radiation.

## Introduction

Microwaves are commonly used in order to aid in the synthesis [1, 2], purification [3, 4], and functionalization [5, 6] of carbon nanotubes (CNTs). However, the behavior of CNTs when exposed to microwaves is not completely understood. Imholt *et al.* [7] observed that when no chemical reagents are present, CNTs heat to temperatures greater than 2000 K and emit infrared, visible, and ultraviolet radiation when exposed to microwaves. The CNTs' strong microwave absorption properties and the emissions of visible radiation suggest that CNTs have the potential for use in electromagnetic shielding and mercury- and electrode-free lighting technologies [8]. Furthermore, it has been shown that exposure to microwaves can cause CNTs' ends to open [9], which could make them prospective candidates for the basis of gas storage technologies.

There has been some controversy concerning the mechanism responsible for the phenomenon. Wadhawan *et al.* [10] attributed the behavior of CNTs in microwave fields to the heating of metal catalyst particles. However, subsequent work [11] has shown that the removal of catalyst particles has no significant effect on the microwave absorption properties of CNTs. Ohmic heating has also been mentioned as a possible explanation for the behavior of CNTs in microwaves fields [8, 12]. Despite the fact that CNTs are theoretically ballistic conductors [13] (meaning that no energy should be dissipated due to electron currents), as-synthesized CNTs typically have defects, which could lead to ohmic heating. Hanson and Patch [14] have shown microwave absorption to increase as the number of defects in CNTs increase, and subsequent experiments performed by Cavness *et al.* [15] have shown that single-walled carbon nanotubes (SWCNTs) which have been functionalized emit radiation of greater intensity when exposed to microwaves as compared to pristine SWCNTs. The behavior of CNTs when exposed to microwaves has also been attributed to the transformation of electromagnetic energy into mechanical vibrations [16]. The model of Ye *et al.* [16] proposes that CNTs undergo heating as the result of transverse parametric resonance due to the polarization of CNTs in the microwave field [8], and experiments performed by Ferguson *et al.* [12] have shown long SWCNTs to emit radiation of greater intensity than short SWCNTs when exposed to microwaves (which could also suggest ohmic heating as the mechanism responsible for the observed heating and radiation emissions).

*scott.williams@angelo.edu

Although microwaves are commonly used to help synthesize, purify, and functionalize CNTs, there have been surprisingly few studies performed to investigate the interactions of CNTs and microwaves at a fundamental level (i.e. at low pressures and in the absence of chemical reagents). The main goals of the experiments described here were to determine how the diameters of multi-walled carbon nanotubes (MWCNTs) effect their behavior in microwave fields, and to determine whether multiple exposures to microwaves results in substantial structural modifications of MWCNTs. Spectra of the radiation emitted by MWCNTs of various diameters when exposed to microwaves were obtained, as well as spectra of the radiation emitted by a single sample of MWCNTs during multiple microwave irradiations. Microwave-induced changes in the structures of MWCNTs were investigated using Raman spectroscopy, which is one of the most sensitive and informative techniques used to characterize disorder in $sp^2$ carbon materials [17]. The D-band (typically found in Raman spectra at a Raman shift of ~1350 cm$^{-1}$) is defect-dependent and is the result of phonons elastically scattering near the K-point of the Brillouin zone of graphite, while the first-order G-band (typically found at a Raman shift of ~1580 cm$^{-1}$) is due to the two-fold degenerate $E_{2g}$ mode at the $\Gamma$-point. The D- to G-band peak-intensity ratio is a commonly-used benchmark that is proportional to the defect densities of $sp^2$ carbon structures.

**Experimental Procedure**

All five of the MWCNT samples used in the experiments were procured from Cheap Tubes (USA). Three of the samples contained MWCNTs with lengths of 10-30 μm and diameters of < 8 nm, 10-20 nm, and 20-30 nm. The other two samples contained MWCNTs with lengths of 10-20 μm and diameters of 30-50 nm and > 50 nm. All five samples had purities of > 95 wt% and ash contents of < 1.5 wt%. Fig. 1 is an SEM image showing the > 50 nm-diameter MWCNTs.

MWCNT samples with masses of 150±0.3 mg were irradiated with microwaves while in a Pyrex test tube at pressures ranging from 4.9×10$^{-5}$ torr to 9.6×10$^{-5}$ torr. The 2.45 GHz microwaves were produced using a magnetron positioned approximately 0.8 cm away from the test tube containing the MWCNTs. The MWCNTs were irradiated with microwaves for durations of 20 seconds. Spectra of the radiation emitted by the MWCNTs while exposed to the microwaves were obtained using an Ocean Optics spectrometer probe, which was placed approximately 1.5 cm away from the bottom of the test tube. The orientations of the magnetron antenna, Ocean Optics probe, and Pyrex tube were the same in all experiments.

Raman spectra were obtained using a Jobin Yvon HORIBA LabRam Raman spectrometer equipped with a 633 nm-wavelength He-Ne laser. Background contributions were subtracted from the spectra using software developed by Candeloro *et al.* [18]. An example of a typical Raman spectrum (without background subtracted) is shown in Fig. 2, and comparisons of the Raman spectra (with background subtracted) of pristine and microwave-irradiated MWCNTs of different diameters are shown in Fig. 3. The spectra shown in Fig. 3 have been normalized with respect to the G-band intensity.

**Results and Discussion**

During the experiments, the MWCNTs continuously emitted visible radiation while they were exposed to microwaves. In addition to the glowing, intermittent flashes of light were also observed, which were likely the result of outgassing. The spectra of the radiation emitted by the MWCNT samples while being irradiated with microwaves is shown in Fig. 4 (with background subtracted). Despite the presence of a few spectral lines (including what appears to be the 656 nm-wavelength line associated with hydrogen), Fig. 4 shows that the spectra of the radiation emitted by the MWCNTs are essentially continuous (and resemble a blackbody radiation spectrum), indicating that the radiation emitted is not primarily the result of some chemical process. Several of the spectra also feature broad peaks (mostly in the infrared region of the spectra) which do not appear to be due to characteristic radiation.

Fig. 4 indicates that, when exposed to microwaves, MWCNTs with larger diameters emit more intense infrared and visible radiation than MWCNTs with smaller diameters (the peak intensity of the radiation produced by the MWCNTs with > 50 nm-diameters is 572% greater than that produced by the MWCNTs with < 8 nm-diameters). The one exception in Fig. 4 is the spectrum associated with the MWCNTs with diameters of 10-20 nm, which is more intense than the spectrum produced using the MWCNTs with diameters of 20-30 nm at essentially all wavelengths and more intense than the spectrum associated with the 30-50 nm-diameter MWCNTs at wavelengths in the range ~600-800 nm. While the exact reason for this is not entirely clear, a second round of experiments showed these results to be reproducible. In general, the MWCNTs with diameters of > 50 nm and 30-50 nm produced the most and second-most intense radiation, respectively, despite the fact that they were slightly shorter in length than the MWCNTs contained in the other samples.

The theoretical work of Shuba *et al.* [19] has shown the polarizing effects of microwave fields to be stronger for MWCNTs with larger diameters than for MWCNTs with smaller diameters of the same length. The same report also showed the polarizability of long MWCNTs to be greater than the polarizability of shorter MWCNTs of the same diameter. When considered in the context of the theory of Shuba *et al.*, the data shown in Fig. 4 and the results of the experimental work of Ferguson *et al.* [12] both suggest that ohmic heating caused by the polarization of the MWCNTs in the microwave field may be responsible for the emissions of visible and infrared radiation from the MWCNTs.

Fig. 5 shows the Raman D- to G-band peak-intensity ratios ($I_D/I_G$) of the MWCNT samples with diameters of > 8 nm, 10-20 nm, 20-30 nm, and 30-50 nm before and after 20 seconds of microwave irradiation. The horizontal error bars represent the ranges of MWCNT diameters present in the samples, with the data points centered at the mean diameter values. The vertical error bars represent the uncertainties in the $I_D/I_G$ values, which were calculated by combining the uncertainties in background subtraction (estimated to be 5%) and statistical uncertainties in quadrature.

In general, Fig. 5 shows that for pristine MWCNTs (as-purchased) $I_D/I_G$ values decrease slightly as MWCNT diameters increase, suggesting that defect densities also decrease with increasing diameters. This is in agreement with the results of Antunes *et al.* [20], who attributed the trend to differences in the defect densities of MWCNT end-caps. A comparison of the $I_D/I_G$ values of the pristine MWCNTs to those of the MWCNT samples that were irradiated indicates that, with the exception of the 30-50 nm-diameter MWCNTs, exposure to microwaves did not significantly affect the defect densities of the MWCNTs. In the case of the MWCNTs with diameters of 30-50 nm, microwave absorption resulted in a 45% increase in the $I_D/I_G$ value. It has been suggested that microwaves can induce currents in residual catalytic metal particles (which often are located at nanotube ends), leading to the tube ends opening [9]. Thus, the change in the $I_D/I_G$ value for the MWCNTs with diameters of 30-50 nm may be due to microwave absorption resulting in the opening of tube ends, which has also been observed by Alvarez-Zauco *et al.* [9].

The spectra of the radiation emitted by a sample containing MWCNTs with diameters of > 50 nm during four 20-second irradiation cycles is shown in Fig. 6 (with background subtracted). The sample was allowed to cool at ambient temperature for approximately 5 minutes between irradiation cycles. Fig. 6 only includes the spectra of emitted radiation with wavelengths in the range 350 nm-614 nm due to detector saturation at wavelengths greater than 614 nm after the first irradiation cycle.

It is evident from a comparison of the spectra that the intensity of the radiation emitted by the MWCNTs while exposed to microwaves increased over the course of the irradiation cycles before reaching a limiting magnitude. These results are in agreement with a previous study involving SWCNTs [15]. Once again, this suggests that the primary mechanism responsible for the MWCNTs' emissions of infrared, visible, and ultraviolet radiation is likely not a chemical reaction. Furthermore, the sustained visible-light emissions from the MWCNTs suggest that they have the potential for use in illumination technology.

Fig. 7 compares the $I_D/I_G$ value for a sample containing pristine MWCNTs with diameters of > 50 nm to $I_D/I_G$ values for MWCNTs with diameters of > 50 nm that have undergone one, two, three, and four 20-second microwave-irradiation cycles. Uncertainties in the $I_D/I_G$ values were calculated by summing statistical uncertainties and the uncertainties in background subtraction (estimated to be 5%) in quadrature.

The data in Fig. 7 does not suggest any stable trend, nor does it suggest any major change in $I_D/I_G$ values as the result of multiple microwave-irradiation cycles. The study performed by Alvarez-Zauco *et al.* yielded similar results [9]. However, the data does show a slight (16%) decrease in $I_D/I_G$ values as the result of repeated exposure to microwaves, indicating a modest decrease in defect densities. This may be the result of the microwaves supplying the MWCNTs with sufficient energy to reorient $sp^3$ bonds (defects) into $sp^2$ hybridization [8]. This overall trend is in contrast with the increase in the $I_D/I_G$ value for the 30-50 nm-diameter MWCNTs seen in Fig. 5. This discrepancy may be due to the fact that Raman spectroscopy and $I_D/I_G$ values in particular have been shown to be sensitive to CNTs' orientations and/or alignments [20].

**Conclusions and Future Work**

A comparison of the radiation emitted by MWCNTs with diameters ranging from < 8 nm to > 50 nm suggested that MWCNTs with larger diameters emit much more intense infrared and visible radiation than MWCNTs with smaller diameters. When these results are considered in the context of the theoretical work of Shuba *et al.* [19], they suggest that the observed behavior of MWCNTs when exposed to microwaves is caused by ohmic heating as the result of the dynamic polarization of the MWCNTs. In general, 20-second exposures to 2.45 GHz microwaves did not have a substantial effect on $I_D/I_G$ values, indicating that irradiation did not significantly impact the MWCNTs' defect densities.

A single sample containing MWCNTs with diameters of > 50 nm was exposed to microwaves during four 20-second irradiation cycles. The intensities of the visible radiation emitted by the MWCNTs during the irradiations increased before reaching a limiting value over the course of the four irradiation cycles. Comparisons of the $I_D/I_G$ values of pristine MWCNTs to those of MWCNTs that had undergone one, two, three, and four irradiation cycles did not indicate a major change in the $I_D/I_G$ values, which indicates that the radiation emissions from MWCNTs exposed to microwaves may be sustainable. These results are also consistent with ohmic heating.

While the data shown here suggest that MWCNTs have the potential for use in lighting technologies, there is still a need for more experimental data in order to gain a better understanding of the microwave-CNT interaction at both fundamental and applied levels. Recently, during the late stages of this paper's preparation, we performed experiments similar to those discussed here which involved a 150 mg sample of $C_{60}$ fullerenes (99.5 wt%), rather than MWCNTs. During exposure to 2.45 GHz microwaves, we observed radiation emissions from the fullerenes that were similar to the radiation emissions from MWCNTs that have been described here (although there was no evidence of outgassing from the fullerenes during irradiation). However, during a subsequent experiment involving a much larger sample (800 mg), no radiation emissions were observed from the fullerenes. Previously, Palstra *et al.* [21] have reported the luminescence of a $C_{60}$ crystal doped with alkali metal when electric currents were passed through it. The spectrum of the radiation emitted by the doped $C_{60}$ crystal in the experiments performed by Palstra *et al.* is very similar to the spectra shown in Fig. 4, suggesting that the same mechanism is responsible for the radiation emissions from both CNTs and fullerenes. Comparisons of the experimental results of Palstra *et al.* [21] (in addition to the theoretical results of Shuba *et al.* [19]) to the experimental results presented here indicate that the mechanism responsible for the emissions of visible and infrared radiation from MWCNTs is likely ohmic heating due to the polarization of the MWCNTs in the microwave field.

## Acknowledgement

The authors wish to thank Winston Layne at the University of Texas at Dallas for his assistance with obtaining the Raman spectra.

Fig. 1. SEM image of sample containing pristine (as-purchased) MWCNTs with diameters of > 50 nm

Fig. 2. Example of a typical MWCNT Raman spectrum (before background subtraction), including the D-band and the G-band, shown at Raman shifts of ~1350 cm$^{-1}$ and ~1580 cm$^{-1}$, respectively

Fig. 3. D- and G-bands of the Raman spectra (normalized with respect to the G-band intensity) of pristine and microwave-irradiated MWCNTs with diameters of < 8 nm, 10-20 nm, 20-30 nm, 30-50 nm, and > 50 nm

Fig. 4. Spectra of the radiation produced by the MWCNTs during 20 seconds of microwave-irradiation

Fig. 5. Raman D- to G-band peak-intensity ratios ($I_D/I_G$) of the MWCNT samples with diameters of > 8 nm, 10-20 nm, 20-30 nm, and 30-50 nm before (black circles) and after (white triangles) 20 seconds of microwave irradiation

Fig. 6. Spectra of the radiation emitted by MWCNTs (diameters > 50 nm) during the first, second, third, and fourth 20-second microwave-irradiation cycles.

Fig. 7. Raman D- to G-band peak-intensity ratios ($I_D/I_G$) for a pristine MWCNT sample (diameters > 50 nm) and samples that have undergone one, two, three, and four microwave-irradiation cycles

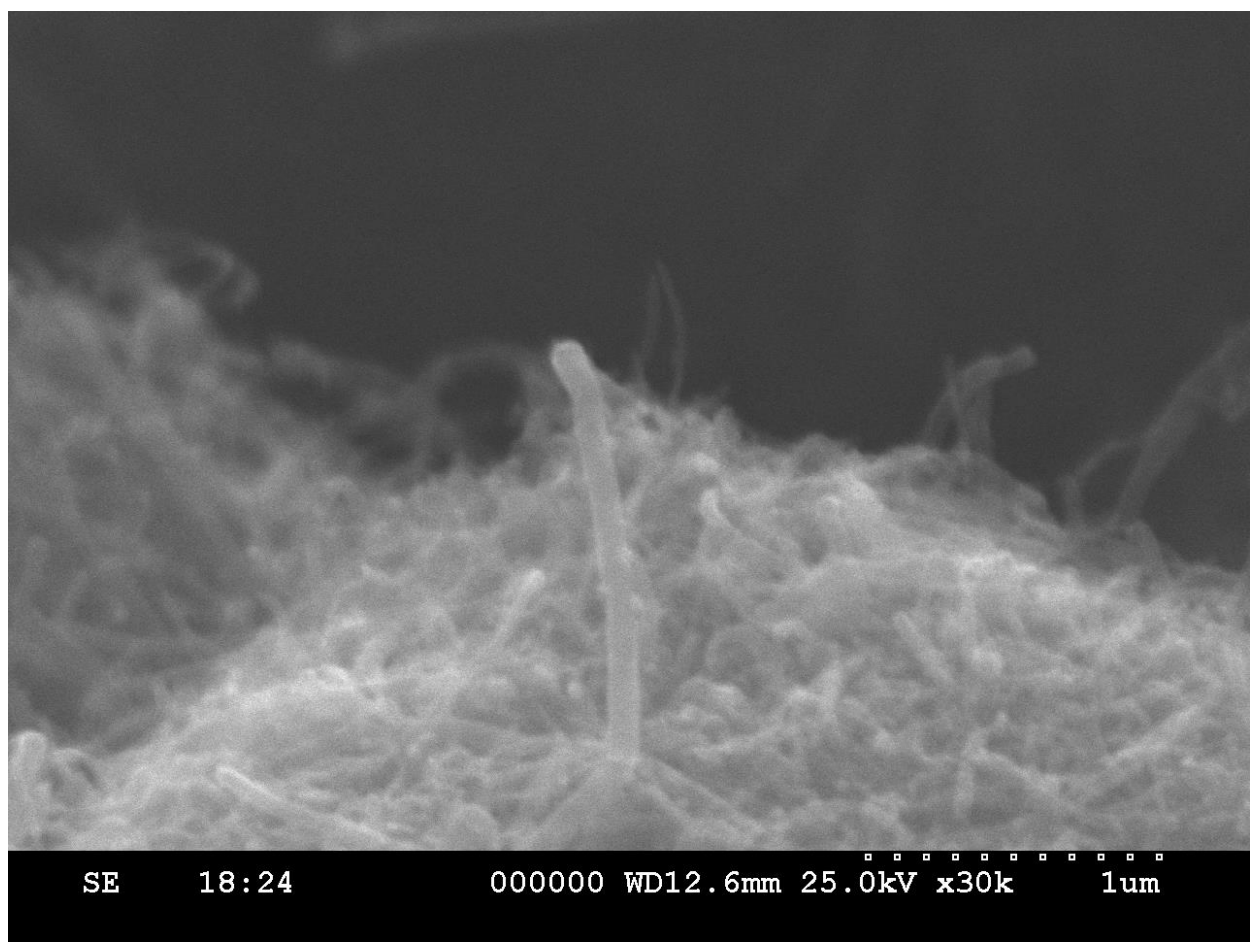

SE    18:24        000000 WD12.6mm 25.0kV x30k      1um

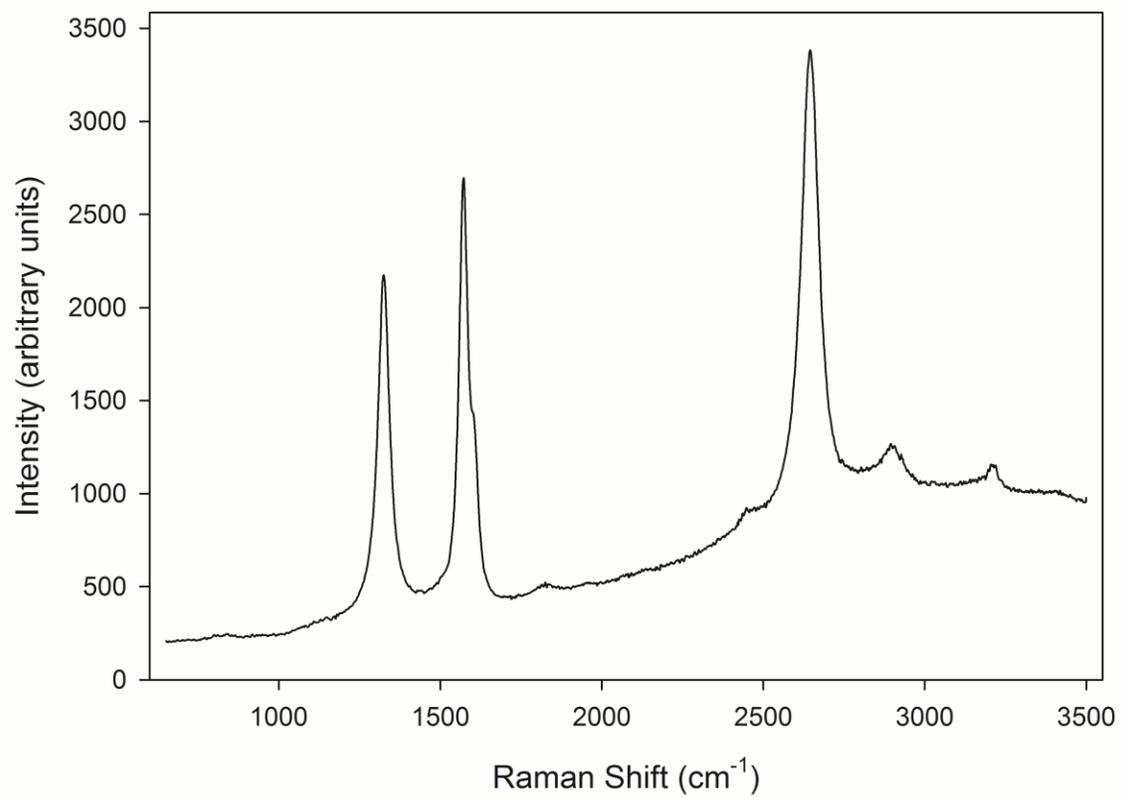

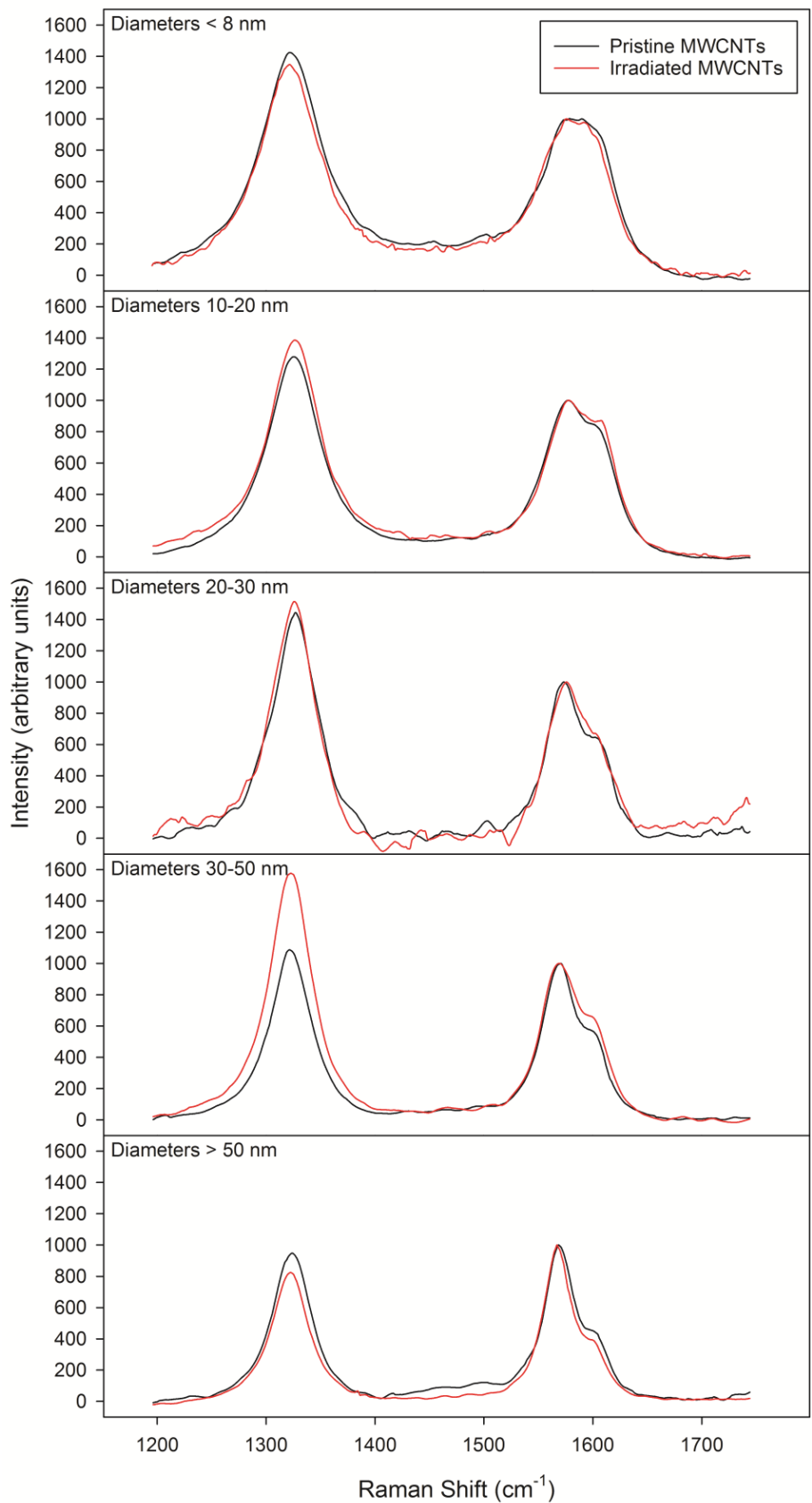

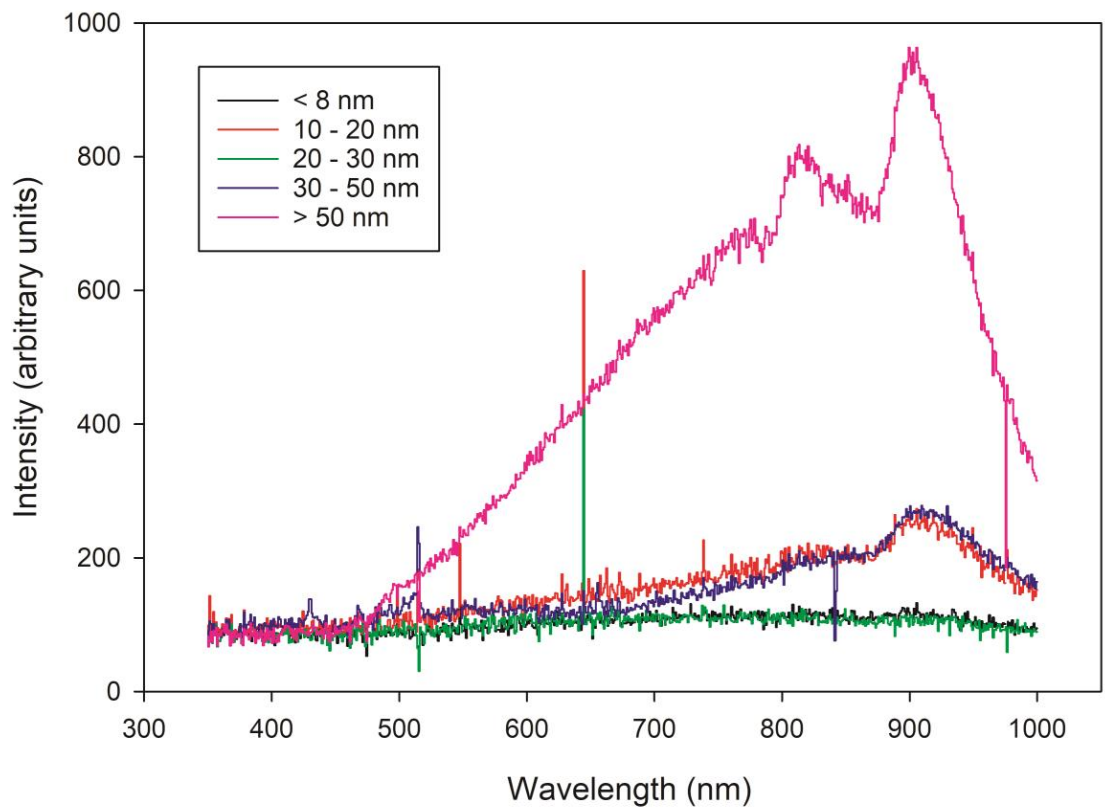

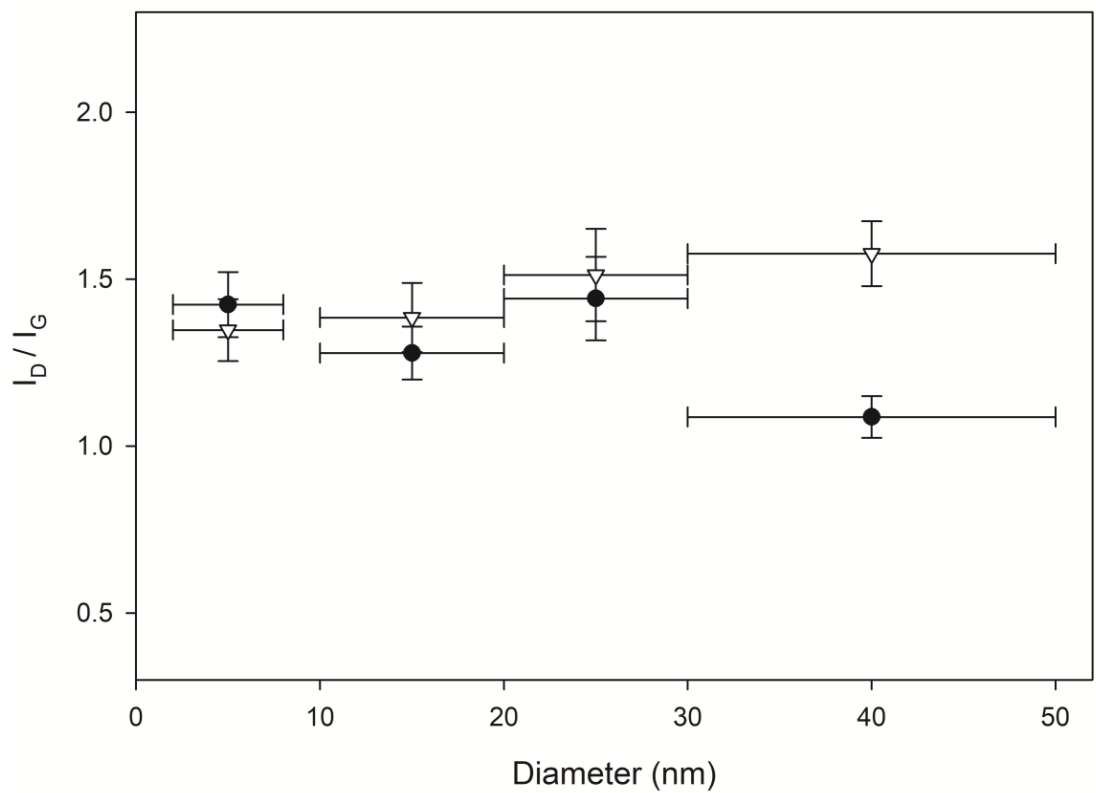

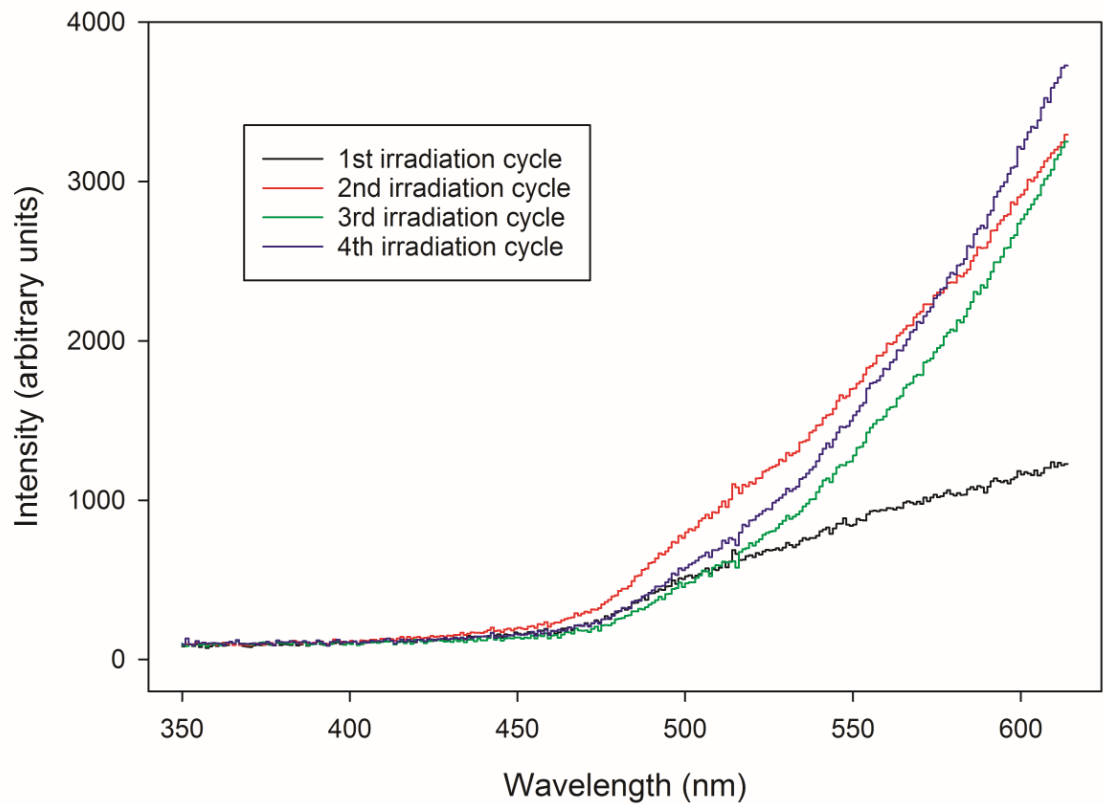

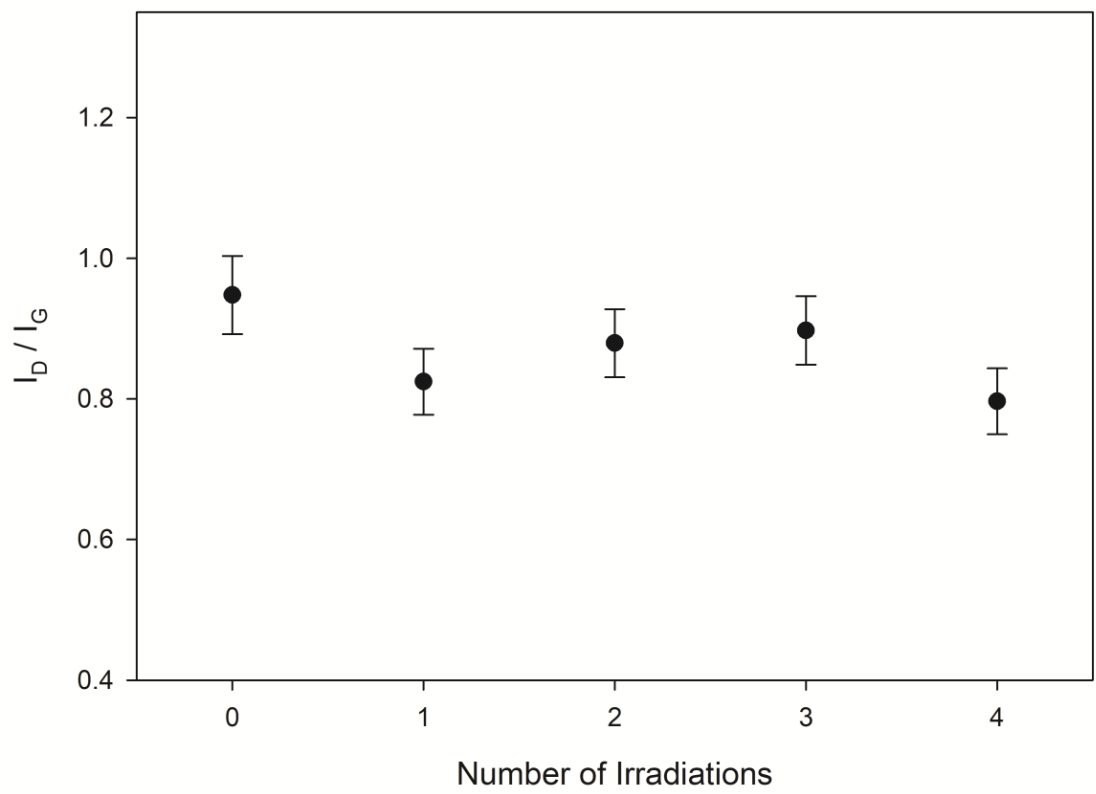